\documentclass[usenatbib]{article}

\usepackage{times,graphicx,amssymb,amsmath,natbib,color}
\usepackage[left=1in, right=1in]{geometry}
\pdfoutput=1
\begin{document}

\newcommand{\mnras}{MNRAS}
\newcommand{\apj}{ApJ}
\newcommand{\nat}{Nat}
\newcommand{\apjl}{ApJL}
\newcommand{\apjs}{ApJS}
\newcommand{\physrep}{Phys.~Rep.}
\newcommand{\aap}{A\&A}
\newcommand{\aaps}{A\&AS}
\newcommand{\araa}{ARA\&A}
\newcommand{\aj}{AJ}
\newcommand{\prd}{PhRvD}
\newcommand{\repprog}{Rep.~Prog.~Phys}

\newcommand{\degrees}{^{\circ}}
\newcommand{\msol}{M_{\rm \odot}}
\newcommand{\Zsol}{Z_{\rm \odot}}
\newcommand{\mjup}{M_{\rm Jup}}
\newcommand{\rjup}{R_{\rm Jup}}
\newcommand{\mearth}{M_{\rm \oplus}}

\newcommand{\todo}[1]{{\color{red}#1}}

\title{Evaluating Galactic Habitability Using High Resolution Cosmological Simulations of Galaxy Formation}
\author{Duncan Forgan $^{1}$\thanks{E-mail:dhf3@st-andrews.ac.uk}, Pratika Dayal $^{2}$, Charles Cockell$^{3}$ and Noam Libeskind$^{4}$} 

\maketitle

\begin{small}
\noindent $^{1}$Scottish Universities Physics Alliance (SUPA), School of Physics and Astronomy, University of St Andrews, KY16 9SS \\
$^{2}$ Institute for Computational Cosmology, Department of Physics, University of Durham, South Road, Durham DH1 3LE, UK\\
$^{3}$ UK Centre for Astrobiology, School of Physics and Astronomy, University of Edinburgh \\
$^{4}$Leibniz-Institute for Astrophysics, Potsdam, An der Sternwarte 16, Potsdam, Germany, 14482
\end{small}

\maketitle

\begin{abstract}

We present the first model that couples high-resolution simulations of the formation of Local Group galaxies with calculations of the galactic habitable zone (GHZ), a region of space which has sufficient metallicity to form terrestrial planets without being subject to hazardous radiation. These simulations allow us to make substantial progress in mapping out the asymmetric three-dimensional GHZ and its time evolution for the Milky Way (MW) and Triangulum (M33) galaxies, as opposed to works that generally assume an azimuthally symmetric GHZ.

Applying typical habitability metrics to MW and M33, we find that while a large number of habitable planets exist as close as a few kiloparsecs from the galactic centre, the probability of individual planetary systems being habitable rises as one approaches the edge of the stellar disc.  Tidal streams and satellite galaxies also appear to be fertile grounds for habitable planet formation.

In short, we find that both galaxies arrive at similar GHZs by different evolutionary paths, as measured by the first and third quartiles of surviving biospheres.  For the Milky Way, this interquartile range begins as a narrow band at large radii, expanding to encompass much of the galaxy at intermediate times before settling at a range of 2-13kpc.  In the case of M33, the opposite behaviour occurs - the initial and final interquartile ranges are quite similar, showing gradual evolution.  This suggests that galaxy assembly history strongly influences the time evolution of the GHZ, which will affect the relative time lag between biospheres in different galactic locations. We end by noting the caveats involved in such studies and demonstrate that high resolution cosmological simulations will play a vital role in understanding habitability on galactic scales, provided that these simulations accurately resolve chemical evolution.


\end{abstract}

\section{Introduction}

\noindent A useful paradigm in astrobiology is ``the habitable zone'', a preferred location or region in space for planets to produce a biosphere \citep{Huang1959,Hart_HZ}.  Most commonly, this term is applied to individual star systems.  Radiative transfer calculations of the atmospheric response of an Earthlike planet to insolation \citep{Kasting_et_al_93,Kopparapu2013,Kopparapu2014} allow the identification of planetary orbits that permit surface liquid water.  This is the principal definition of "the habitable zone", and we henceforth refer to it as the stellar habitable zone (SHZ).

As a consequence of stellar evolution, the SHZ is time-dependent.  The increasing luminosity of ageing stars tends to push the habitable zone further outwards \citep{Rushby2013} - we can therefore define a continuously habitable zone (CSHZ) as the region which remains inside the SHZ for the entirety of the star's hydrogen fusion phase, usually referred to as the Main Sequence.

Binary systems have especially non-trivial SHZs, as the contribution to insolation from both sources, as well as perturbations of the planet's orbit, must be accounted for \citep{Eggl2012,Kaltenegger2013, Haghighipour2013,Mason2013,Forgan2012,Forgan2014,Cuntz2014}.  The interaction of the two stars also influences the planet's high energy radiation environment \citep{Johnstone2015,Zuluaga2015}. 

Exomoons may also provide sites for biospheres, and their SHZs are complicated even further by the effects of tidal heating \citep{Heller2013b, Forgan_moon1}, frequent eclipses \citep{Heller2012} and by infrared radiation from the host planet \citep{Heller2013,Forgan2014a}.  These SHZs implicitly assume that magnetic fields can shield the moon from high energy radiation, which bears its own caveats \citep{Heller2013a}.

Despite these complications and uncertainties, the SHZ remains one of the principal tools used by astrobiologists to assess the potential of exoplanet systems for producing habitable worlds.  To date,  29 planets with physical radii less than 2.5 times the Earth's radius, with orbits inside their local SHZ\footnote{Planetary Habitability Laboratory http://phl.upr.edu/projects/habitable-exoplanets-catalog accessed 11th March 2015}.  This is of course only the first step towards determining if these worlds are indeed habitable, as their atmospheric chemistry and tectonic activity will play a crucial role (see \citealt{Guedel2014} for a review).

Somewhat less discussed is the habitable zone at galactic scales, as was first proposed by \citet{Gonzalez2001}.  In principle, we should be able to identify Galactic Habitable Zones (GHZs), regions of galaxies that are more amenable to planet formation and the growth of a biosphere, while being relatively unmolested by hazardous events such as nearby supernovae, gamma ray bursts, close stellar encounters etc.  

This implies a balance.  Terrestrial planet formation requires the formation of chemical elements up to iron and beyond.  Astronomers use the concept of \emph{metallicity} to describe the fraction of elements with higher atomic number than helium, so in these terms a minimum metallicity must be generated locally for terrestrial planet formation to proceed. Metallicity generation occurs in the cores of stars, so metallicity generation requires star formation and stellar death to release the requisite elements into the interstellar medium, so that they might be incorporated into planets.  Equally, stellar death often results in violent mechanical and radiative events such as supernovae (SNe).  The local supernova rate must be reasonably low to prevent large doses of hazardous radiation sterilising terrestrial planets.   In analogy with the SHZ, the GHZ does not guarantee that planets inside it are habitable, but merely indicates more favourable conditions for habitability.

The above is a quite broad explanation of the primary factors that define the GHZ - in practice, the GHZ is troublesome to elucidate, as it depends on a host of secondary factors that are difficult to model, and as such there is a great deal of disagreement over the shape and evolution of the Milky Way's GHZ.

\citet{Gonzalez2001}'s initial study outlines the key constraints on the Milky Way's GHZ - the metallicity gradient, and the supernova rate distribution.  This analysis results in the ``classic'' picture of the GHZ as an annulus in the thin disc, which moves outwards in time as metallicity is generated in the Galaxy.  They assume a minimum metallicity, of approximately half that of solar, is required for terrestrial planet formation.  In their picture, increasing metallicity increases the chance for planet formation (but their calculations of radiogenic heating show that elemental composition is as important as simply finding elements more massive than hydrogen and helium).

\citet{GHZ} would improve on this analysis with a more sophisticated model of the distribution of terrestrial planets in space and time \citep{Line_planets}, in particular employing the latest statistical data on the probability of planets given stellar metallicity, rather than assuming a simple metallicity threshold for planet formation.  This work indicated that the Milky Way's GHZ is an annulus between 7 and 9 kiloparsecs from the Galactic Centre, with the Sun close to the centre of this zone at approximately 8 kiloparsecs.

\citet{Prantzos2007} was highly critical of these approaches, and noted the uncertainties that then surrounded the planet-metallicity relation.  In particular, the planet-metallicity relation applied more strictly to giant planets, especially given the observational biases placed upon the data at that time.  Indeed, it seemed that high metallicity could damage terrestrial planet formation through the migration of Hot Jupiters from the outer planetary system to very close orbits. Given this, and the uncertainties in the metallicity gradient, \citet{Prantzos2007} proposed that the GHZ could encompass the entire Milky Way disc! 

In light of this, later works have been more critical of the GHZ limits they produce.  One of the most sophisticated models of the Milky Way's GHZ is that of \citet{Gowanlock2011}, which attempts to map the GHZ by modelling individual star systems.  As a result, they are better placed to describe habitability above and below the Galactic midplane, as well as the properties of individual planetary systems.  They give a much more detailed account of both Type I and Type II supernovae, as well as tidal locking of planets to their parent star.  

In recent years, other galaxies have had their GHZs mapped, in particular elliptical galaxies. \citet{Suthar2012}, working principally from metallicity constraints, suggested that the elliptical M87, and the dwarf galaxy M32 could possess relatively broad GHZs.  Later works \citep{Carigi2013} used one dimensional galaxy chemical evolution models to map the GHZ of Andromeda (M31), which incorporated more realistic physics, such as radial mixing \citep{Spitoni2014} to improve the metallicity gradient.

It is therefore clear that each galaxy's GHZ depends critically on its evolutionary history, in particular on its accretion history.  Understanding this history is a crucial goal of galaxy formation theory, which aims to explain the observed evolution of galaxy properties as a function of time.

According to the standard paradigm, Dark Matter (hereafter DM) collapses to form halos with complex baryonic physics governing galaxy formation within these potential wells \citep{White1978}. However, the process of galaxy formation is much more complicated, with DM halos growing both through anisotropic accretion and mergers and galaxy formation proceeding by the combined action of clumpy and anisotropic gas accretion and mergers of dwarf galaxies.  These processes result in clumpy and disky galaxies whose anisotropic shapes are only now beginning to be captured by the latest hydrodynamic cosmological simulations (e.g. \citealt{Stinson2006, libeskind2010,Kobayashi2011, knebe2011,Pilkington2012,Scannapieco2012,Vogelsberger2014, Graziani2015}).

Most studies of the GHZ to date assume a relatively steady, monotonic growth of the Milky Way from an infall model, or in the case of \citet{Gowanlock2011}, the Milky Way's stellar distribution is initialised in its present day configuration, and azimuthal symmetry is still assumed (their study is restricted to a 1$^\circ$ sector of the Galaxy to boost spatial resolution).

In this work, we take a different approach to GHZ modelling.  We analyse cosmological simulations of galaxy formation, which were designed to explore the formation history of the Local Group, whose principal components are the Milky Way, Andromeda (M31) and Triangulum (M33) galaxies.  By analysing the habitability of star particles in the simulation, we are able to investigate non-axisymmetric habitable zones, and their non-trivial evolution with redshift\footnote{Astronomers often use \emph{redshift}, $z$, in place of time.  This is more convenient, as $z$ is a measurable quantity of a galaxy's spectrum, and is a fundamental variable of cosmological models of the Universe's expansion history.  A redshift $z=\inf$ refers to the beginning of the Universe at the Big Bang, and $z=0$ is the present day, with the relationship between $z$ and the Universe's age being a non-linear function of various cosmological parameters}.

\section{Method}

\subsection{Cosmological Simulations}

We use the {\tt CLUES}\footnote{http://www.clues-project.org} simulations for the calculations presented in this work. We briefly summarize the simulations here and the interested reader is referred to \citet{libeskind2010}, \citet{knebe2010}, \citet{libeskind2011} and \citet{knebe2011} for complete details on the central galaxies, their discs and the properties of their $z=0$ substructure population, as well as details regarding the simulation initial conditions, gas-dynamics and star formation. 

The simulations used in this study were run with the PMTree-Smoothed Particle Hydrodynamics (SPH) code \texttt{GADGET2} \citep{springel2005} in a cosmological box of size $64 h^{-1}$ comoving Mpc ($\rm {cMpc}$)\footnote{comoving units are typically used in cosmology to ``factor out'' the expansion of the Universe}. The runs employed the standard cosmological model of the Universe, referred to as $\Lambda$CDM due to the presence of both cold dark matter and dark energy as the principal contributions to the Universe's energy density.  This defines standard $\Lambda$CDM initial conditions for the particles and parameters for the relative contributions of each component of the Universe's energy density, which are taken from the WMAP3 data release \citep{spergel2007} such that $\Omega_m = 0.24$, $\Omega_{b} =0.042$, $\Omega_{\Lambda} = 0.76$, $h = 0.73$, $\sigma_8 = 0.73$ and $n=0.95$ (the reader is referred to \citealt{Peacock1999} for definitions of these terms).  

The $z = 0$ density field is described by a reconstruction that uses observations of objects in the local universe as constraints. Initial conditions are then selected using the Hoffman-Ribak method \citep{hoffman-ribak1991}. These constrained initial conditions force the $z = 0$ linear scales to resemble the local universe; non-linear scales are unconstrained. To obtain a reliable local group (LG) comprising the Milky Way (MW) and Andromeda (M31), around 200 low-resolution ($256^3$ particles) constrained runs were performed until a suitable candidate (defined in terms of the mass, relative distance and velocity of the MW and M31 haloes) was found. Once selected, particles in a sphere of radius $2h^{-1} cMpc$ at the $z = 0$ location of this region were then traced back to the initial condition (at $z=100$) and replaced with higher resolution particles (equivalent to having $4096^3$ particles spanning the full cosmological box) using the prescriptions given by \citet{Klypin2001}. This refining results in producing two objects resembling the MW and M31 in the high resolution region, each of which contains about $10^6$ particles within their virial radii at $z=0$. Outside of this region, the simulation box was populated with lower resolution (i.e higher mass) particles to mimic the correct large scale environment of the local group. The DM and baryons (i.e. gas and star particles) in the refined region have masses of $2.1 \times 10^5h^{-1}\msol$ and $4.4 \times 10^4h^{-1}\msol$, respectively.

\begin{figure}
\begin{center}
\includegraphics[scale=0.3]{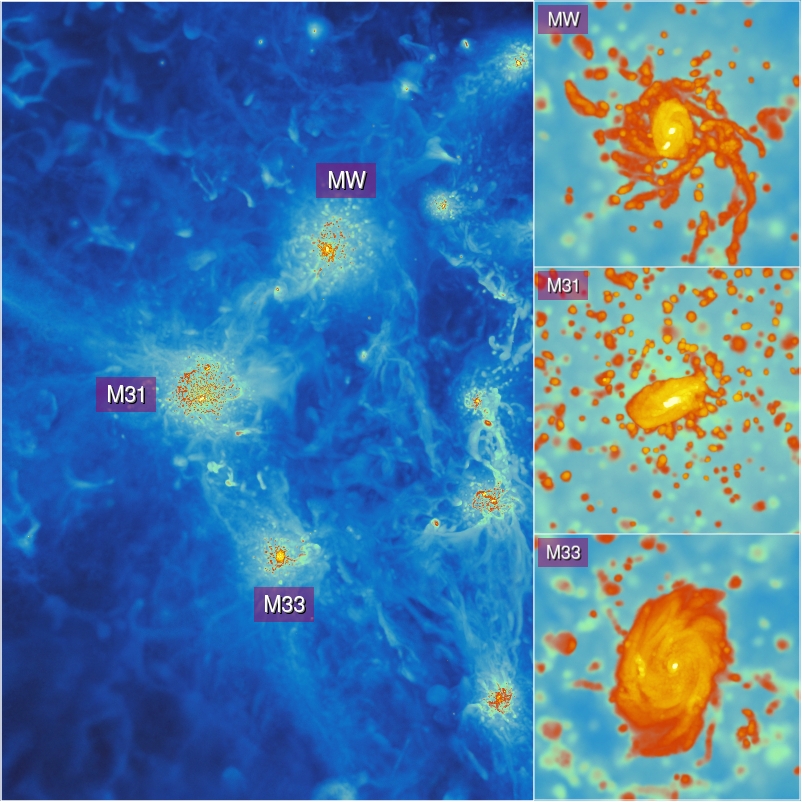}
\caption{The gas distribution of the Local Group in the \texttt{CLUES} simulations on large scales (left picture, about $2 Mpc/h$ across, viewed from a distance of $3.3 Mpc/h$) and the gas disks of the three main galaxies (right panels, about $50 kpc/h$ across, from a distance of $250 kpc/h$).  For the zoomed pictures the colour mapping is shifted to higher densities (factor $10^0.5$) in order to enhance the spiral arm features of the gas disks.  From the \texttt{CLUES} project, image courtesy of K. Riebe. \label{fig:simulation}}
\end{center}
\end{figure}

Around 500 processors and 1.5 terabytes of memory were used to generate the initial conditions for these simulations. The high-resolution N-Body plus gas (SPH) simulations used 500 MPI processors with a typical computing time of the order of 1 million CPU hours, yielding in excess of 6.1 terabytes of data \citep{Gottloeber2010}.  The resulting gas distribution of the simulated Local Group can be seen in Figure \ref{fig:simulation}.

As for the baryonic modelling, the initial mass function (IMF) used is Salpeter between $0.1-100{\rm M_\odot}$.  This is less than ideal for our purposes, as a Salpeter IMF over-estimates the proportion of low mass stars in a given population, and as a result can distort the calculated supernova rates (cf \citealt{Gibson1997}).  However, our calculations must remain consistent with the simulation's own properties.  The simulation employs the feedback rules of \cite{springel-hernquist2003b}: hot ambient gas and cold gas clouds in pressure equilibrium form the two components of the interstellar medium (ISM). Gas properties are calculated assuming a uniform but evolving UV background generated by QSOs and AGNs \citep{haardt-madau1996}. Metal line (or molecular) dependent cooling below $10^{4}{\rm K}$ is ignored. Star formation is treated stochastically, choosing model parameters that reproduce the Kennicutt law for spiral galaxies \citep{kennicutt1983, kennicutt1998}.   This empirical relation gives the star formation rate surface density as a powerlaw of the surface density of molecular gas (or in some cases the surface density of cold molecular gas, which produces a slightly different powerlaw exponent).

Each gas particle was allowed to undergo two episodes of star formation, each time spawning a star particle of half the original mass (i.e. $2.2\times10^{4}~h^{-1}{\rm M_{\odot}}$).  The star particles have an effective radius of 150 pc, defining a resolution limit for our habitability calculations.  The instantaneous recycling approximation is assumed: cold gas cloud formation (by thermal instability), star formation, the evaporation of gas clouds, and the heating of ambient gas by supernova driven winds all occur at the same instant. We assume kinetic feedback in the form of winds driven by stellar explosions.  Assuming instantaneous recycling limits our ability to make habitability arguments based on the NCHOPS elements, as we cannot model the production of various chemical species over stellar lifetimes. Instead, we must make more general arguments as to the availability of elements above hydrogen and helium in the Periodic Table.

Haloes and subhaloes are identified using the publicly available Amiga Halo Finder \citep[AHF;][]{knollmann-knebe2009}. AHF uses an adaptive grid to find local over-densities; after calculating the gravitational potential, particles that are bound to the potential are assigned to the halo. 

We identify all the particles within the virial radius of MW and M31 at $z=0$ and follow them back in time in order to locate the LG progenitors at high-$z$. If bound to a structure, we identify the corresponding halo: each of these halos was then considered a progenitor of the local group at high-$z$ and used in our calculations, as explained in what follows (see \citealt{Dayal2012,Dayal2013} for complete details of the halo identification procedure).

\subsection{Calculating Habitability}

\noindent The cosmological simulations deliver a list of ``star'' particles $\{i\}$ for each snapshot.  Each particle has a total mass $M_i = 3.1612 \times 10^4 \msol$, equivalent to a moderately sized star cluster, so we should consider each particle as consisting of an ensemble of stars.  

Each particle has an effective radius of around 500 parsecs.  Supernovae at distances of around 8 parsecs or less could produce enough ionising flux at the surface of a terrestrial planet to affect its ozone layer and disrupt the biosphere \citep{Gehrels2003}.  As such, we do not expect neighbouring star particles to produce sufficiently strong supernovae or ionising flux to influence each other's habitability, and as such we can consider the habitability of each particle independently.  We note that in reality, star clusters will dissolve and disperse on timescales of a few Myr, and diffuse azimuthally throughout the galaxy on timescales of an orbital period, i.e. a few 100 Myr (see e.g. \citealt{MoyanoLoyola2015}).  The CLUES simulations cannot model this diffusion, so our calculations do not capture the consequences, e.g. the diffusion of supernova progenitors into the vicinity of other star particles.

Each particle has an associated star formation rate $S_i$, and all share the same initial mass function (IMF), i.e. the same distribution of stellar masses.  The cosmological simulation assumes a Salpeter IMF, and therefore for consistency we also assume it here:

\begin{equation}
P(M_*)dM_* \propto M_*^{-\alpha},
\end{equation}

\noindent where $M_*$ is star mass, and the constant $\alpha=2.35$.  By specifying a minimum and maximum mass (0.1 and 100 $\msol$ respectively), integrating a normalised version of this expression fixes the number of stars in each star particle, as well as the proportions of stars of a given mass, which we use to calculate Type Ia and Type II supernova (SNe) rates.  We are required to assume a Salpeter IMF, but this function overestimates the true proportion of low mass stars which can affect SNe rates by a factor of a few for a given stellar population.

Type II supernovae progenitors are stars with masses above 8 $\msol$, therefore we calculate the fraction of stars that are formed above this critical mass, and multiply this by the star formation rate to obtain a Type II SNe rate.  This approach does not take into account the time delay between star formation and supernova.  The longest lived stars that produce Type II SNe are the lowest mass, and they remain on the Main Sequence for approximately 40 million years.

Type Ia SNe progenitors are white dwarfs, with masses below the Chandrasekhar limit of 1.4 $\msol$.  However, not all stars below this mass will become Type Ia SNe - the process is believed to require the presence of extra material for the white dwarf to accrete, such as a companion star.  As white dwarf progenitors span the 0.08 -8 $\msol$ mass range, and it is estimated that 1\% of the white dwarf population ignites as a Type 1a SNe \citep{Pritchet2008}, we therefore assume that the probability of any individual star below 8 $\msol$ exploding as a Type 1a SN is 0.01.

Again, we assume that the Type Ia SNe rate is directly related to the production rate of progenitors, and does not incorporate a time delay due to stellar evolution, which will be many billions of years for the lowest mass objects. We return to this no-delay issue in the Discussion.   The above prescription fixes the ratio of Type II to Type 1a SNe to $\sim 2$, which is slightly lower than the measured ratios of $3-5$ for disk galaxies (cf \citealt{Mannucci2007}).

Star particles that experience combined SNe rates that are too high will experience a reduced probability that the habitable planets they host will possess surviving biospheres (see later).

The simulation also delivers metallicities for each star particle.  These metallicities allow us to calculate two probabilities related to planets:

\begin{enumerate}
\item The probability that terrestrial planets may form in the SHZ, $P_{planet,t}$
\item The probability that close-in giant planets may form, $P_{planet,g}$
\end{enumerate}

\noindent We wish to compare with azimuthally symmetric studies, so we therefore follow \citet{Gowanlock2011} by suggesting that for a system to produce habitable worlds, it should produce terrestrial planets in the SHZ without producing close-in giant planets.  Giant planets cannot be formed \emph{in situ} close to the star; instead, they are formed at larger distances and migrate inwards.  We assume that such events are destructive to habitable terrestrial planets ``in the way'' of giant planet migration, but we note that simulations indicate that giant planet migration may improve the odds of terrestrial planet formation in the SHZ in some cases \citep{Raymond2005, Fogg2007}.

We assume terrestrial planet formation occurs when the metallicity $Z$ rises above a threshold, following \citet{Prantzos2007}:

\begin{equation}
P_{planet,t} = \begin{cases}
0.4 & \frac{Z}{\Zsol}>0.1 \\
0.0 & \frac{Z}{\Zsol}\leq 0.1 
\end{cases}
\end{equation}

This is in line with observations that show the correlation between host star metallicity and the presence of a planet is much weaker below Neptune masses \citep{Fischer2005,Buchhave2012,Wang2014,Schlaufman2015}.  The exact value of the boundary we use here is not tailored to the latest Kepler results, but it does produce terrestrial planets for a large fraction of star systems, which generally speaking is thought to be the case given the Kepler dataset \citep{Petigura2013}.  Above Neptune masses, the metallicity correlation is stronger, so we assume that the probability of giant planet formation is governed by:

\begin{equation}
P_{planet,g} = \begin{cases}
0.03 \times 10^{log \frac{Z}{\Zsol}} & \frac{Z}{\Zsol}>1 \\
0.03 & \frac{Z}{\Zsol} \leq 1
\end{cases}
\end{equation}

Therefore, the probability of a terrestrial planet being habitable is $(1-P_{planet,g})$, and the number of habitable planets hosted by a star particle is

\begin{equation}
N_{hab} =N_{planet}(1-P_{planet,g}) = N_*P_{planet,t}(1-P_{planet,g})
\end{equation}

Habitable planets that exist in hazardous environments will be sterilised.  We can therefore propose a probability of survival, related to the local supernovae rate $SNR$:

\begin{equation}
P_{survive} = 1 - \frac{SNR}{2SNR_\odot}
\end{equation}

\noindent We propose this \emph{ansatz} to allow the probability of survival to be at a maximum when the local supernovae rate is zero, and to be zero if the SNR is twice the solar rate.  This is similar to many previous studies \citep{GHZ,Carigi2013}, although in some cases a step function is used, as opposed to a linear approximation.  As we cannot track individual stellar positions, we cannot use \citet{Gowanlock2011}'s more detailed approach of tracking individual stellar distances to nearby supernovae, and hence we are forced to use this \emph{ansatz}, which is clearly oversimplified.  This is a weak point in our analysis.  While our results are not sensitive qualitatively to on the above criterion, we should expect some quantitative effects on our calculations, but even a more appropriate, detailed calculation is still subject to general uncertainty in how SNe can sterilise habitable planets (see Discussion).

Once calculated, this survival probability can be multiplied by the number of habitable planets to give the number of planets that remain in clement conditions, $N_{survive}$:

\begin{equation}
N_{survive} = N_{hab}P_{survive}
\end{equation}

We do not place age constraints on habitability, unlike other studies, which demand a minimum of 4 Gyr of clement conditions to produce complex life forms.  The pros and cons of such constraints are addressed in the Discussion.

\section{Results}

For brevity, we analyse only two of the Local Group objects - the Milky Way, and M33.  The object representing Andromeda (M31) was omitted as it did not form a well-defined disc - this can be seen in Fig 3 of \citet{Libeskind2013}\footnote{This highlights an ongoing issue with cosmological simulations with gas dynamics, often referred to as the ``angular momentum problem'' (see e.g. \citealt{Piontek2011}).  This issue is likely to be linked to how the simulations inject energy from supernovae into the interstellar gas \citep{Stinson2006, Zavala2008,Scannapieco2009,Stinson2010}.}

\subsection{M33}



\begin{figure*}
\begin{center}$\begin{array}{ll}
\includegraphics[scale=0.45]{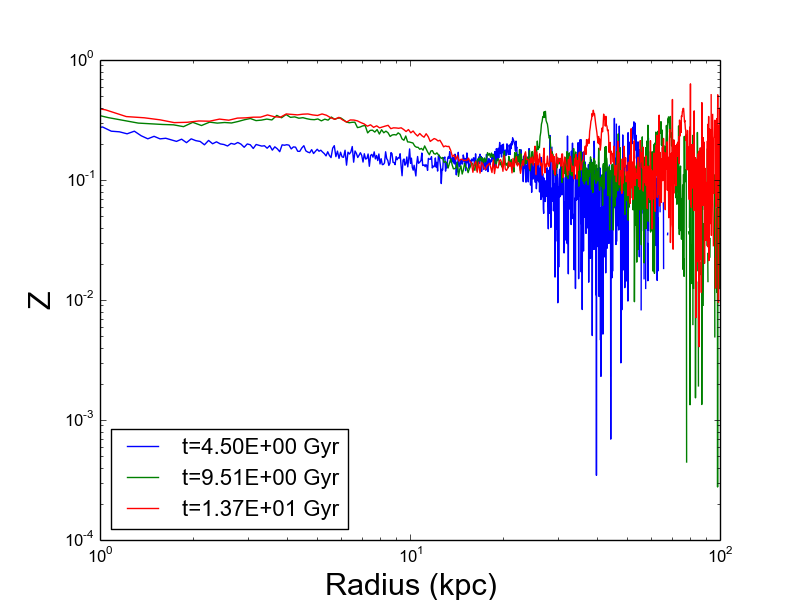} &
\includegraphics[scale=0.45]{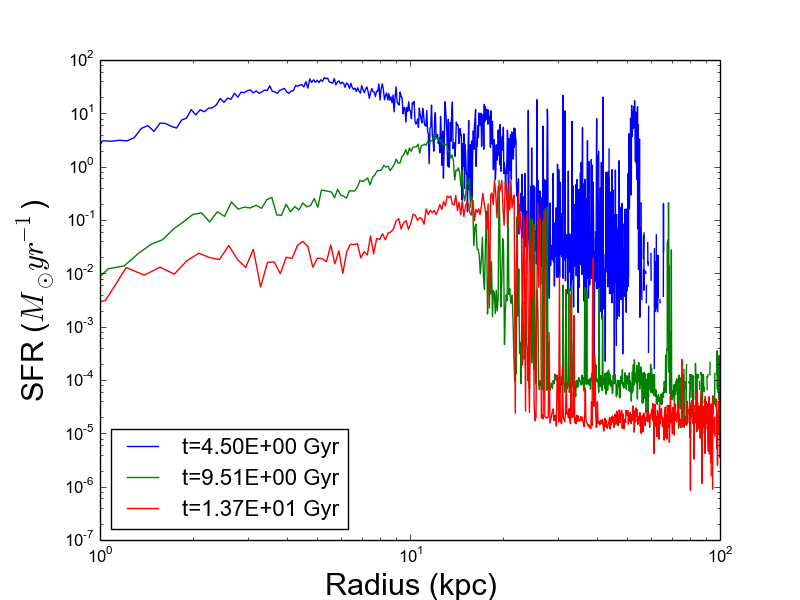} \\
\includegraphics[scale=0.45]{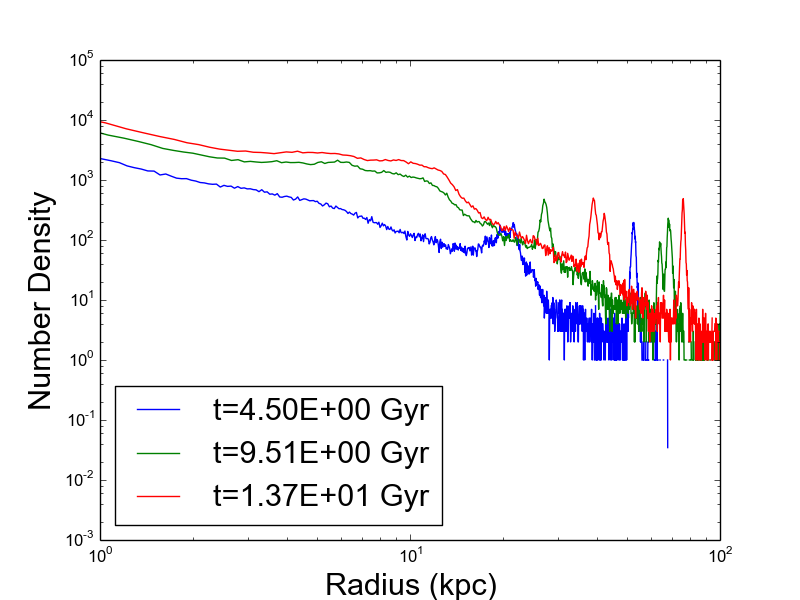} &
\includegraphics[scale=0.45]{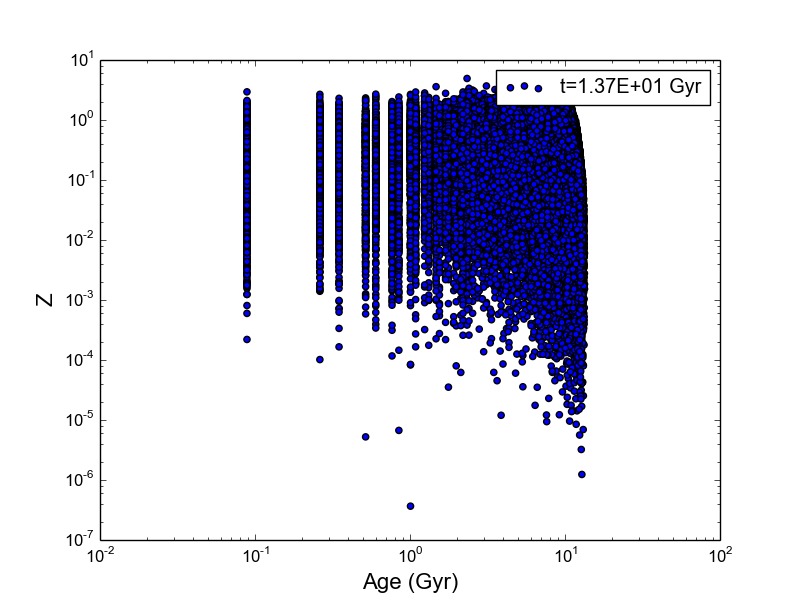} \\
\end{array}$
\caption{The physical properties of the M33 galaxy. We plot the local average stellar metallicity (top left), the local average star formation rate (top right) and the number density of stars (bottom left) at three instances in the simulation, and the age-metallicity relation for all stars in the simulation (bottom right) at $z=0$. \label{fig:M33_properties}}
\end{center}
\end{figure*}

\noindent Figure \ref{fig:M33_properties} shows the evolution of metallicity, star formation rate and number density of stars in the M33 simulation.   The main stellar disc truncates at around 10kpc, with a broadly exponential surface density profile.  The disc is sufficiently metal rich at early times to surpass the threshold for terrestrial planet formation in the inner regions.  However, significant star formation at early times suppresses habitability, with the disc more or less fully formed by $t=9.5$ Gyr.  The age-metallicity relation has a large amount of scatter, which is to be expected as the metals are fully entrained in the gas, and not allowed to diffuse (see e.g. \citealt{Pilkington2012}).

\begin{figure*}
\begin{center}$\begin{array}{ll}
\includegraphics[scale=0.45]{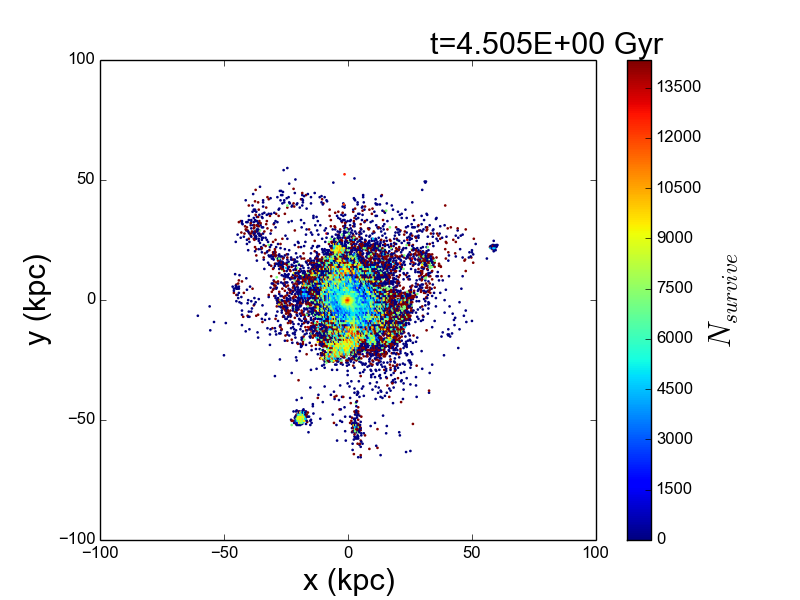} &
\includegraphics[scale=0.45]{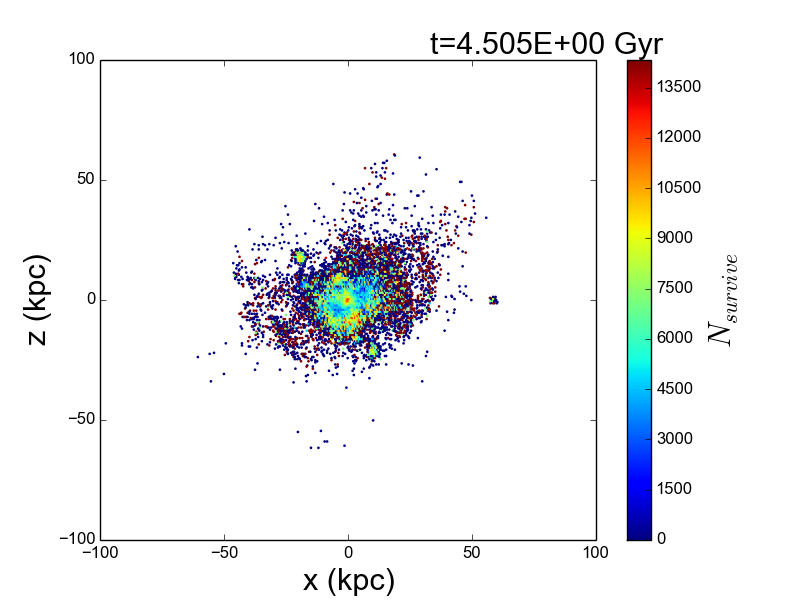} \\
\includegraphics[scale=0.45]{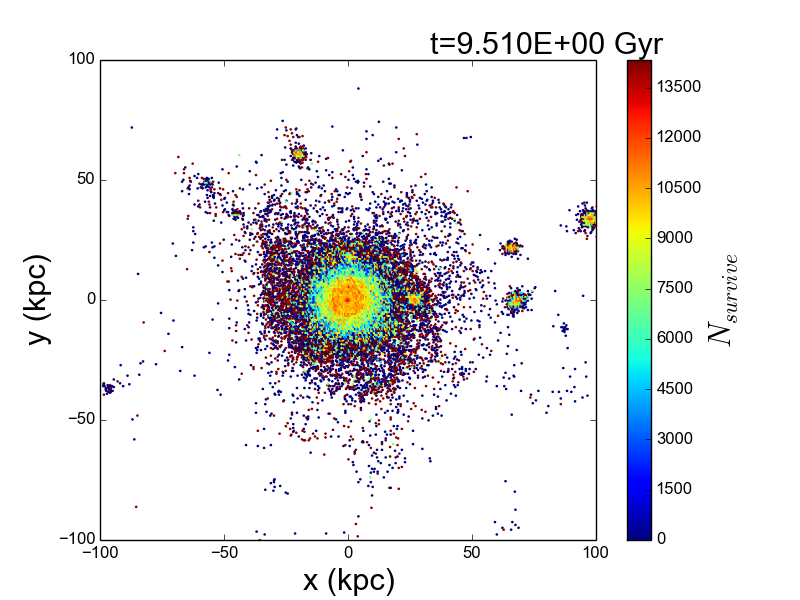} &
\includegraphics[scale=0.45]{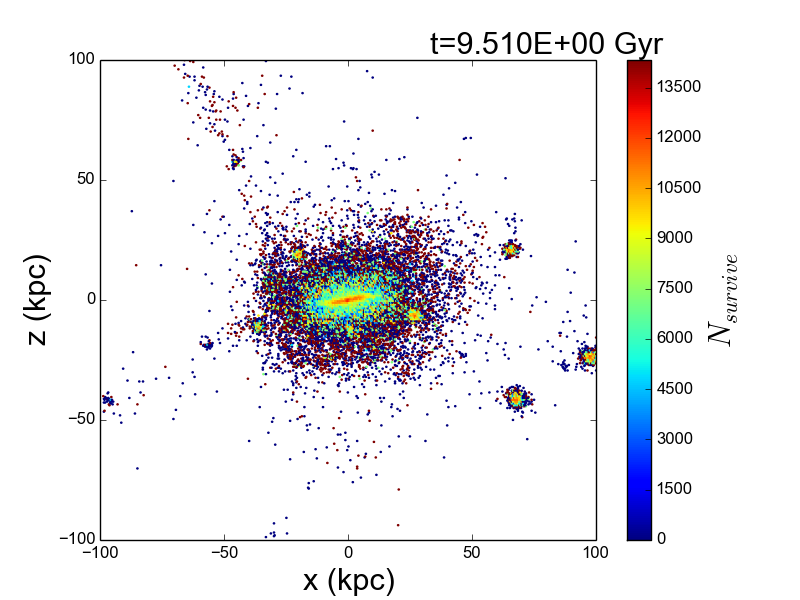} \\
\includegraphics[scale=0.45]{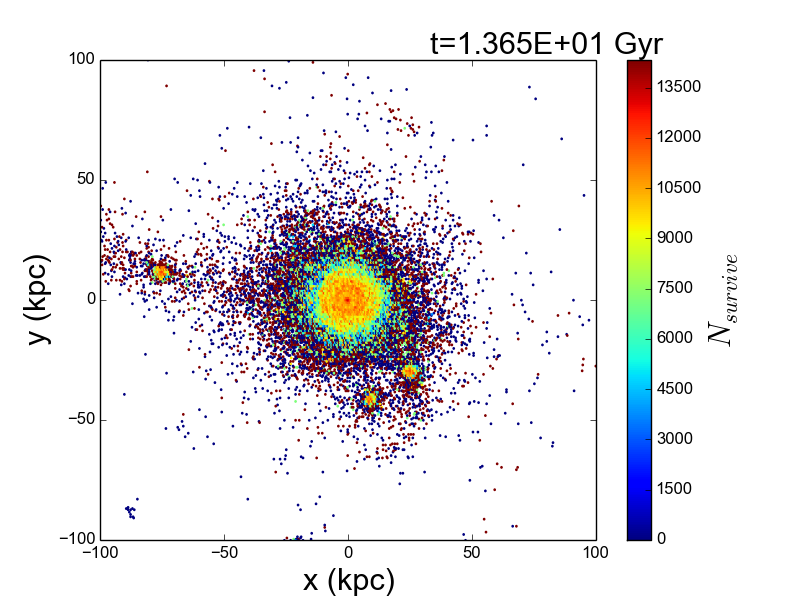} &
\includegraphics[scale=0.45]{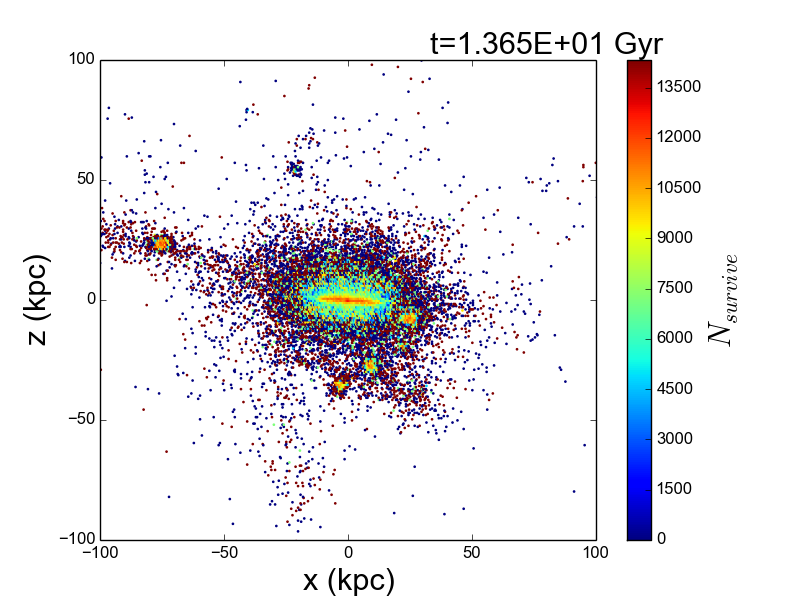} \\
\end{array}$
\caption{The evolution of $N_{survive}$ in the M33 galaxy.  The 2D binned value of $N_{survive}$ is shown for the $x-y$ plane (left) and $x-z$ plane (right).  \label{fig:M33_2D_nsurvive}}
\end{center}
\end{figure*}

\noindent Figure \ref{fig:M33_2D_nsurvive} shows how $N_{survive}$ evolves over the course of the simulation.  Each panel represents a 2D binning to give the average value of $N_{survive}$ (bins with $N_{survive}=0$ are not plotted).  Note that the simulations have been rotated so that the $z$ axis is aligned with the system's global angular momentum vector.

The early snapshots (top panels of Figure \ref{fig:M33_2D_nsurvive}) show indications of two-armed spiral structure, but remain poorly resolved.  A well-defined disc is formed around 4 Gyr from the present day (middle panels of Figure \ref{fig:M33_2D_nsurvive}), but it is clear that the GHZ is very non-axisymmetric before disc formation.  Initially, the most habitable regions are to be found in clumps orbiting greater than 10 kpc from the centre of the Galaxy, and this persists to the present day (bottom panels).  In particular, satellite galaxies in particular appear to have broad GHZs in this analysis.

\begin{figure*}
\begin{center}$\begin{array}{ll}
\includegraphics[scale=0.45]{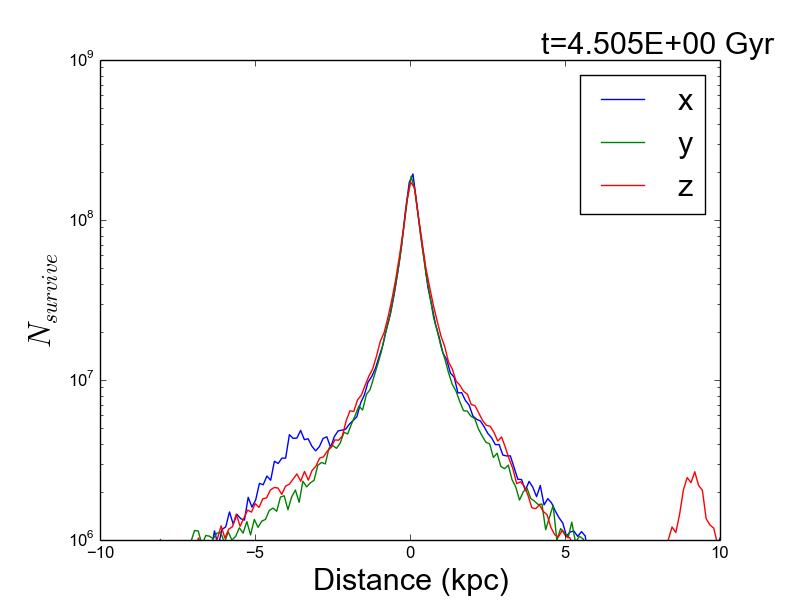} &
\includegraphics[scale=0.45]{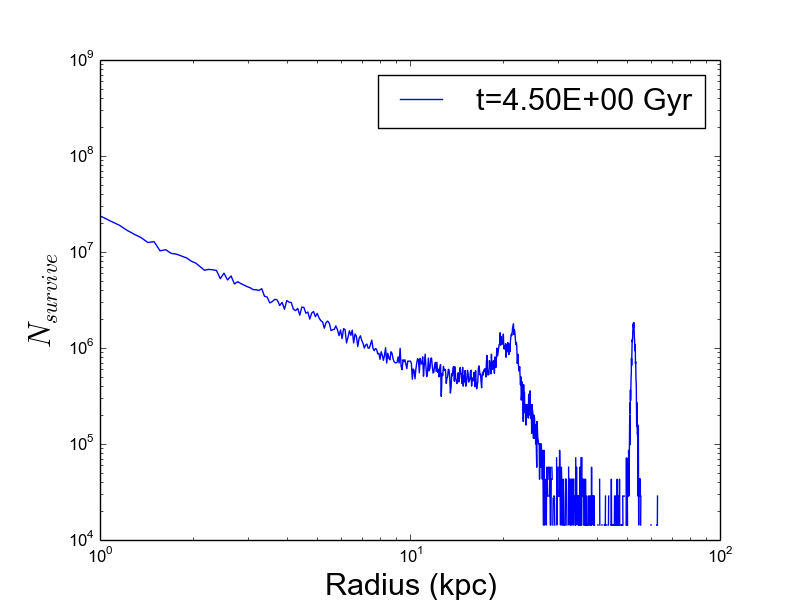} \\
\includegraphics[scale=0.45]{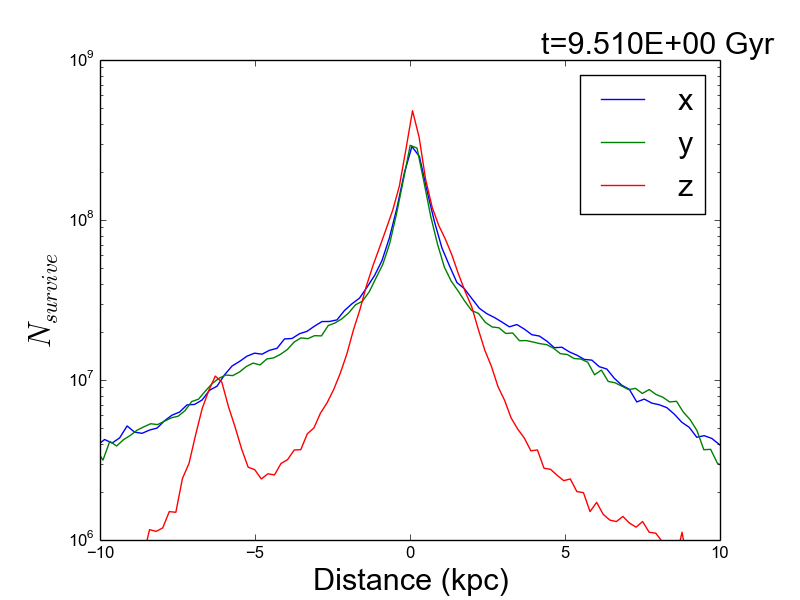} &
\includegraphics[scale=0.45]{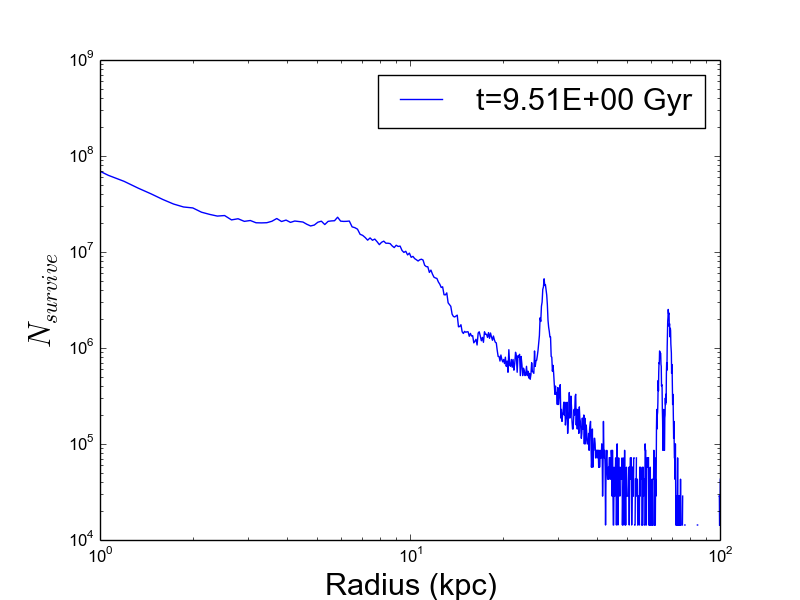} \\
\includegraphics[scale=0.45]{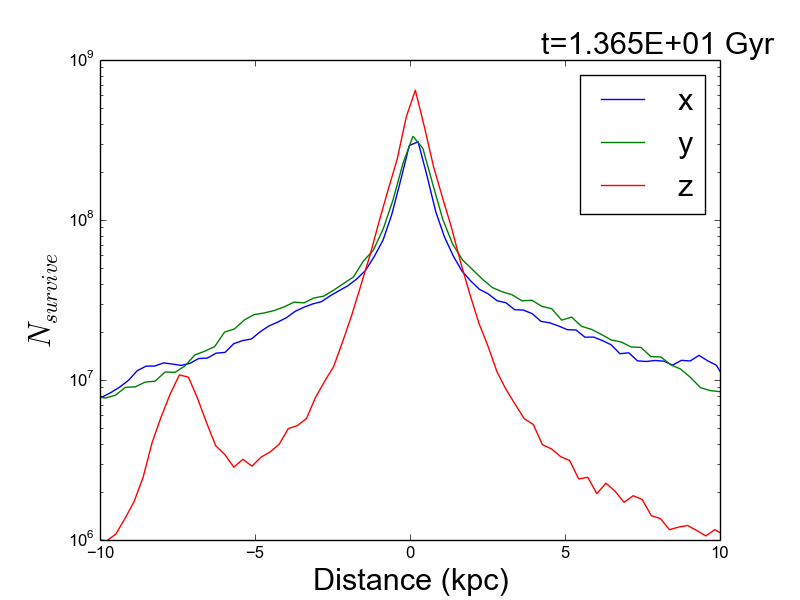} &
\includegraphics[scale=0.45]{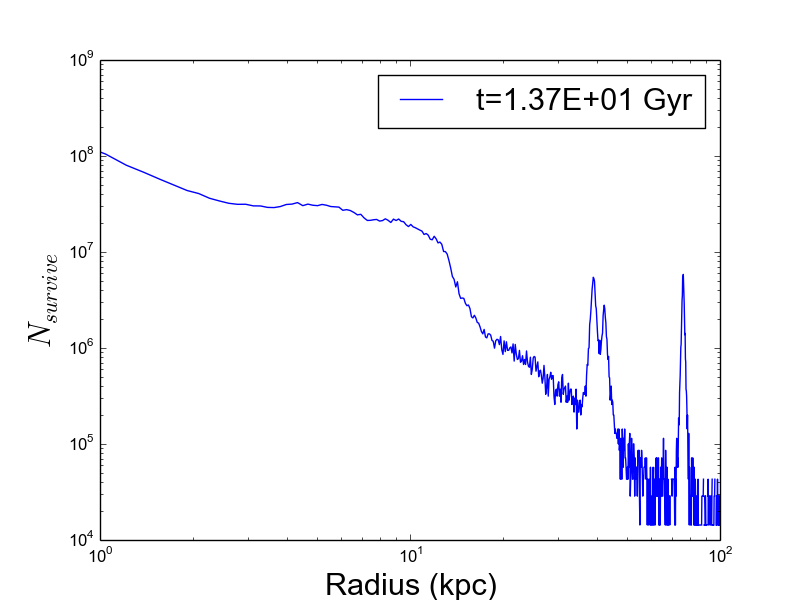} \\
\end{array}$
\caption{Left: The axial distributions of surviving planets in the disc of the M33 galaxy.  In each snapshot, the number of surviving planets is binned along the x,y and z axes.  The strong peak in the $z$-curve between $-10$ and $-5$ kpc is due to the presence of a satellite orbiting above the galactic plane. Right: the radial distribution of surviving planets. \label{fig:M33_axial_nsurvive}}
\end{center}
\end{figure*}

As the disc forms, the habitability of the inner regions increases, but a halo of good habitability remains at the disc's edges.  The radial profiles of M33 (right panels of Figure \ref{fig:M33_axial_nsurvive}) show that large numbers of planets remain unsterilised down to distances of 1 kpc from the centre.  Presumably at distances within 1kpc, the ionising radiation produced by the galaxy's supermassive black hole (not modelled in this analysis) is likely to render this region inhospitable. 

However, we do see that at regions between around 4 and 10kpc, the number of surviving planets increases slightly over the linear trend (in log-log space).  Beyond the disc's edge, spikes in survivability belie the presence of satellites at these galactocentric radii.

As we are analysing 3D data, we have the advantage of comparing the radial habitability of M33 with its habitability along the $x$, $y$ and $z$ axes (left panel of Figure \ref{fig:M33_axial_nsurvive}).  Initially, the simulation shows habitable planets forming in an azimuthally symmetric configuration.  Over time, the formation of the disc narrows the range of habitability in $z$, and the $x$ and $y$ profiles are quite similar (but not identical).  As satellites and streams merge into the disc, we can see asymmetries grow and disappear in the profiles.    As we reach the present day, the GHZ becomes more uniform, but the simulations indicate that such uniformity is a recent phenomenon.  Note that our vertical density distributions are quite similar to those shown by \citet{Juric2008a}, suggesting that these galaxies follow exponential profiles both radially and vertically.


\subsection{Milky Way}

\begin{figure*}
\begin{center}$\begin{array}{ll}
\includegraphics[scale=0.45]{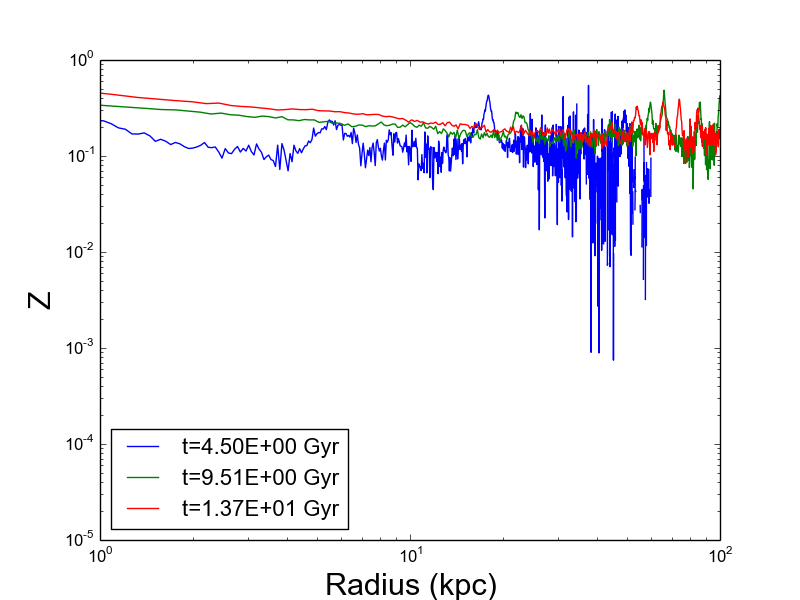} &
\includegraphics[scale=0.45]{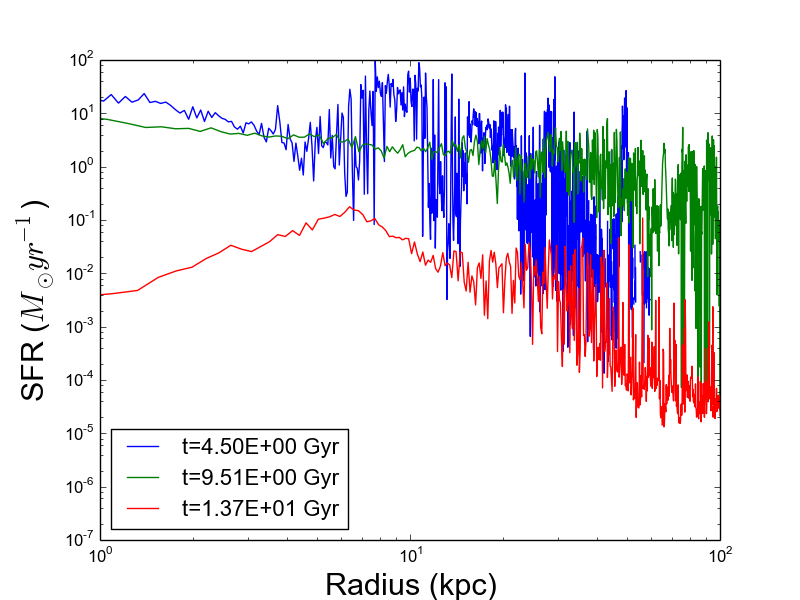} \\
\includegraphics[scale=0.45]{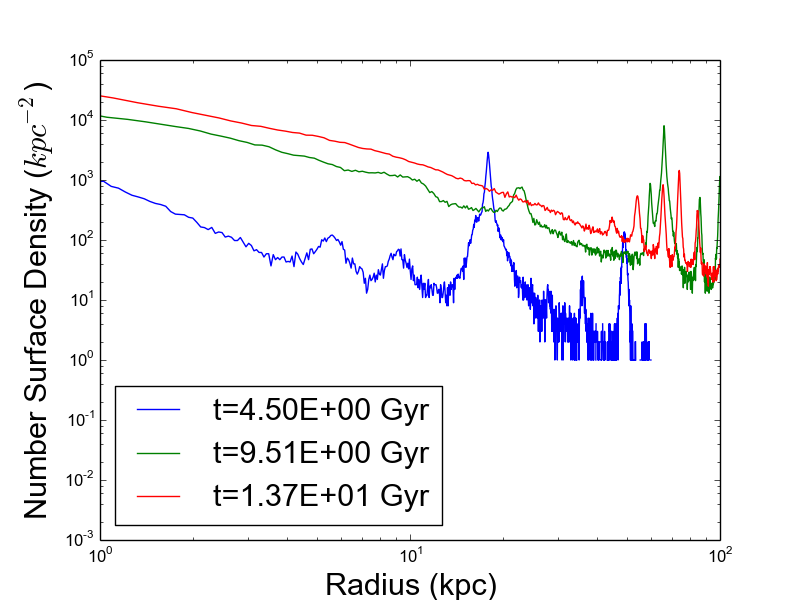} &
\includegraphics[scale=0.45]{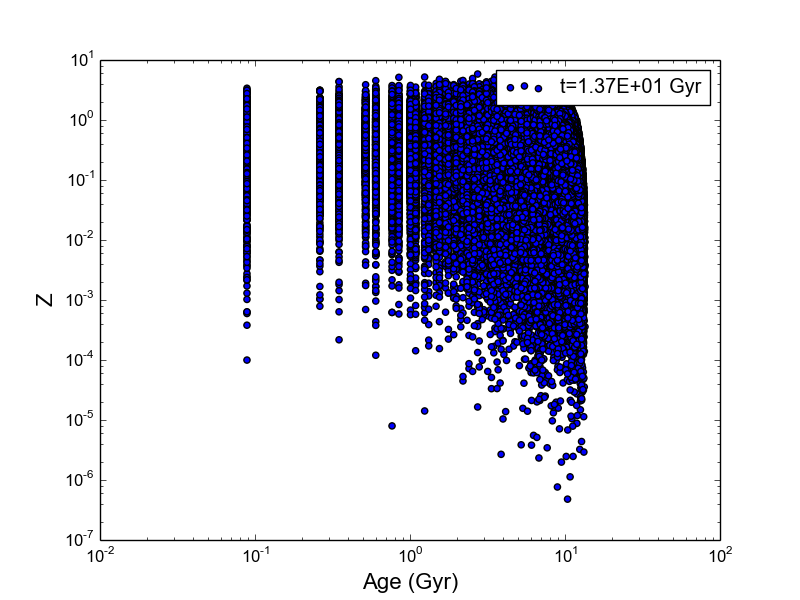} \\
\end{array}$
\caption{As Figure \ref{fig:M33_properties}, but for the Milky Way \label{fig:MW_properties}}
\end{center}
\end{figure*}

\begin{figure*}
\begin{center}$\begin{array}{ll}
\includegraphics[scale=0.45]{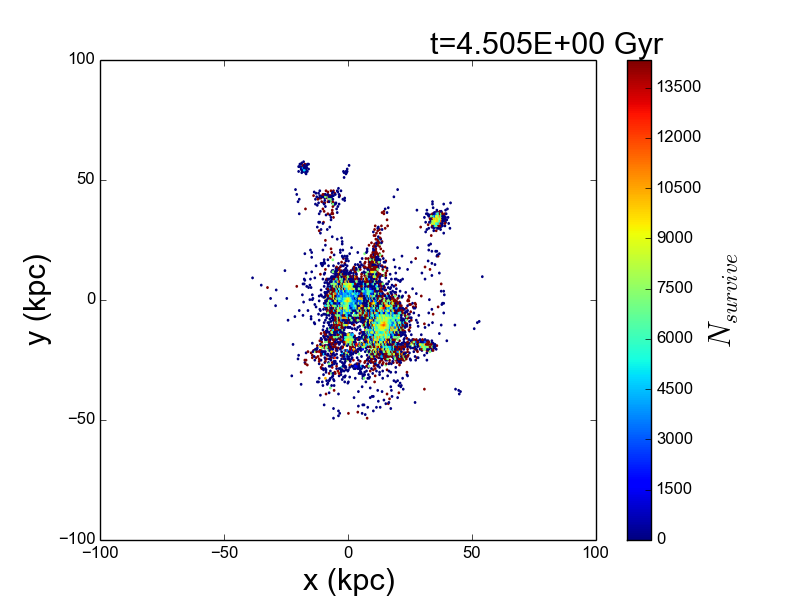} &
\includegraphics[scale=0.45]{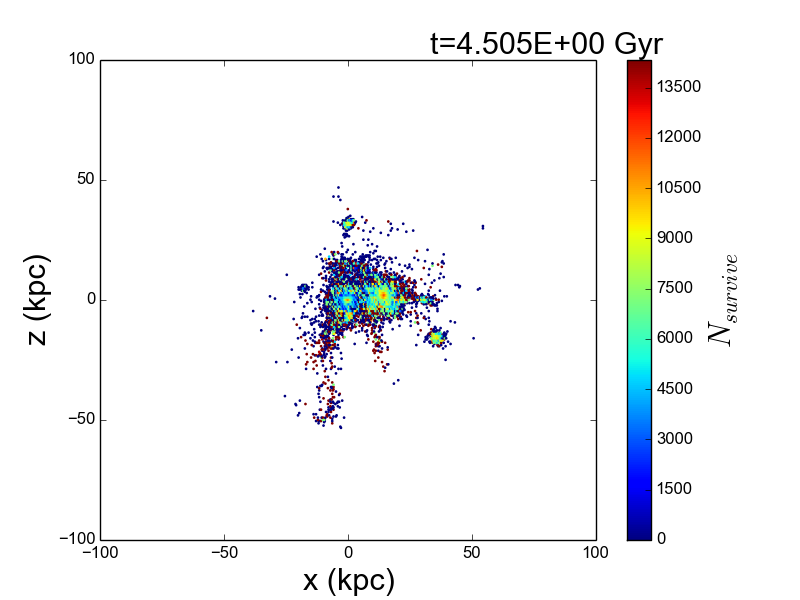} \\
\includegraphics[scale=0.45]{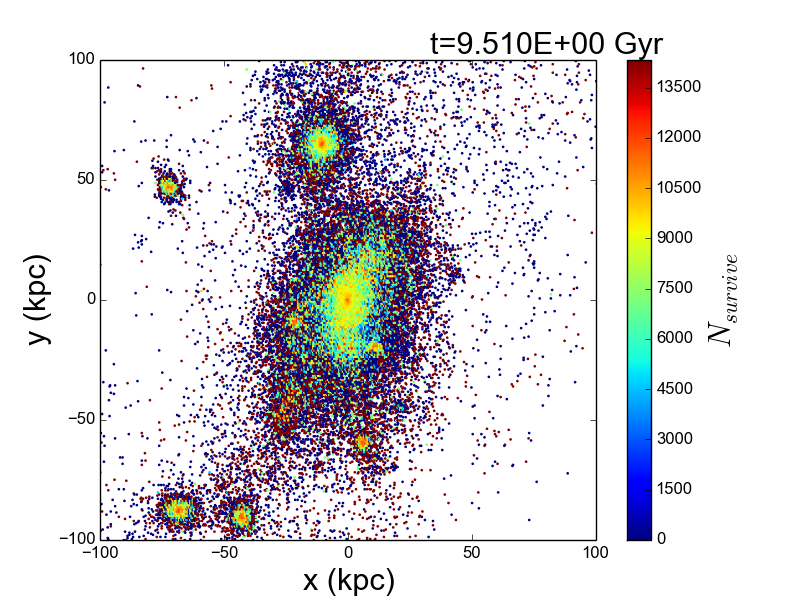} &
\includegraphics[scale=0.45]{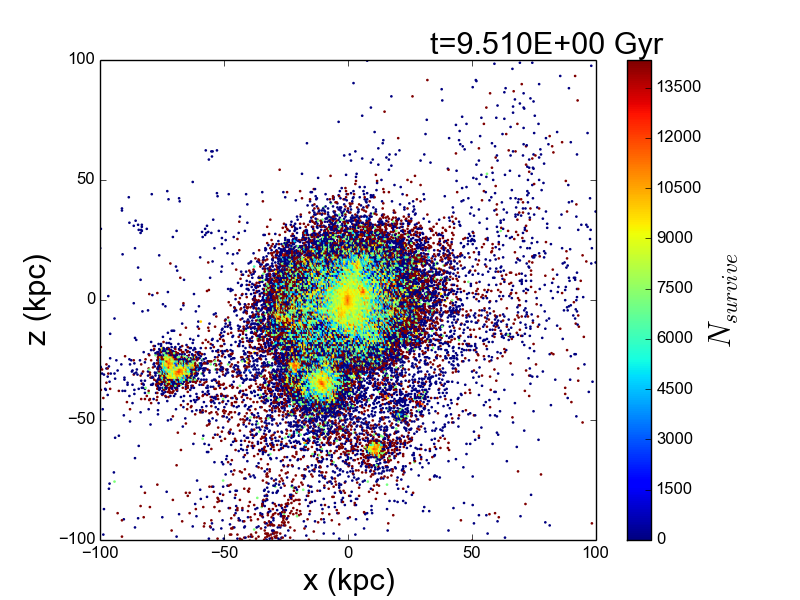} \\
\includegraphics[scale=0.45]{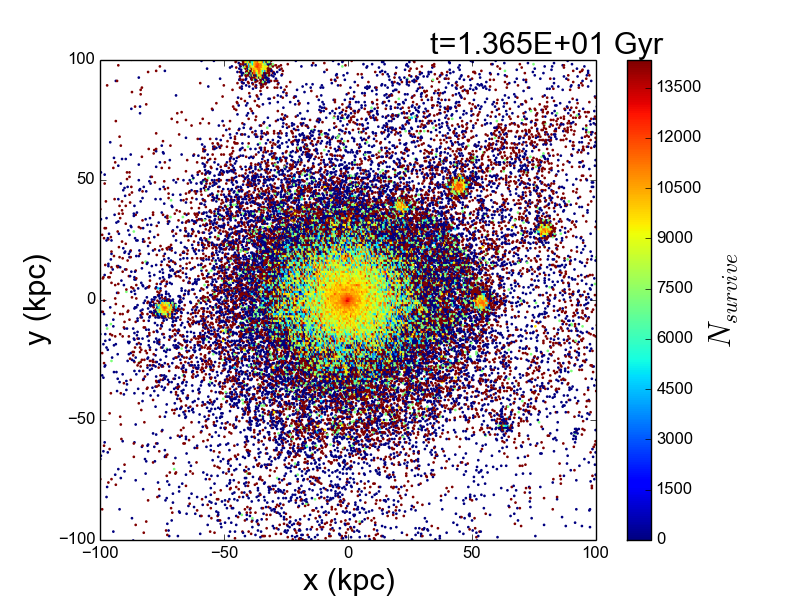} &
\includegraphics[scale=0.45]{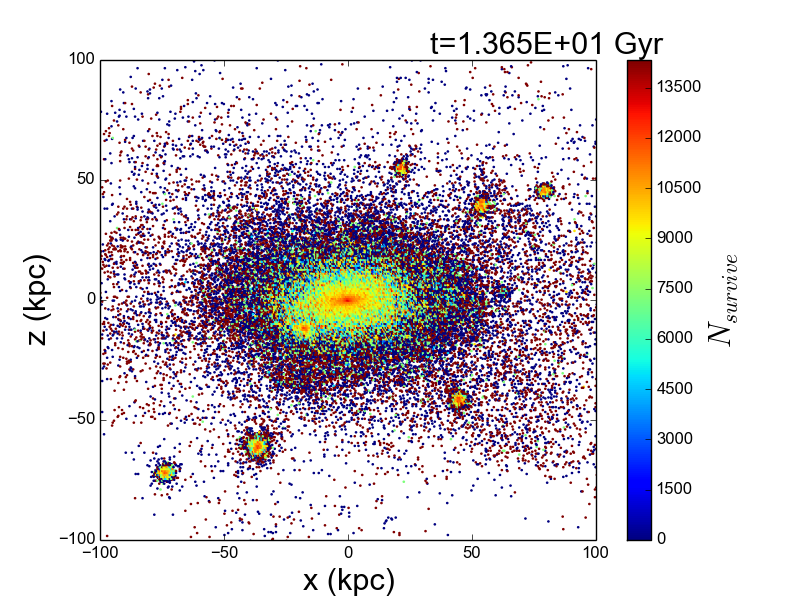} \\
\end{array}$
\caption{As Figure \ref{fig:M33_2D_nsurvive}, but for the Milky Way. \label{fig:MW_2D_nsurvive}}
\end{center}
\end{figure*}

\noindent We now consider the other galaxy in our simulation data, which is selected as it partially resembles the Milky Way. Figure \ref{fig:MW_properties} again shows the galaxy's physical properties, which suggests a more punctuated, chaotic formation history compared to M33.  As can be seen also in Figure \ref{fig:MW_2D_nsurvive}, this galaxy does not form a well-defined disc, but exhibits a more triaxial structure.  It has accreted more material ``off-axis'' from its angular momentum vector, as the flow of material from streams and satellites has been more isotropic. This accretion history is hinted at by the galaxy's large number of satellites, even at the present day, and the satellites in general contain more star particles.  The total number of star particles at the present day is 545,349 which is nearly double that of M33.  Star formation rates remain relatively high even at late times, reducing the number of habitable planets in the outer disc, and as a result favouring locations beyond the disc, especially its larger satellites, for surviving biospheres until much later times.

\begin{figure*}
\begin{center}$\begin{array}{ll}
\includegraphics[scale=0.45]{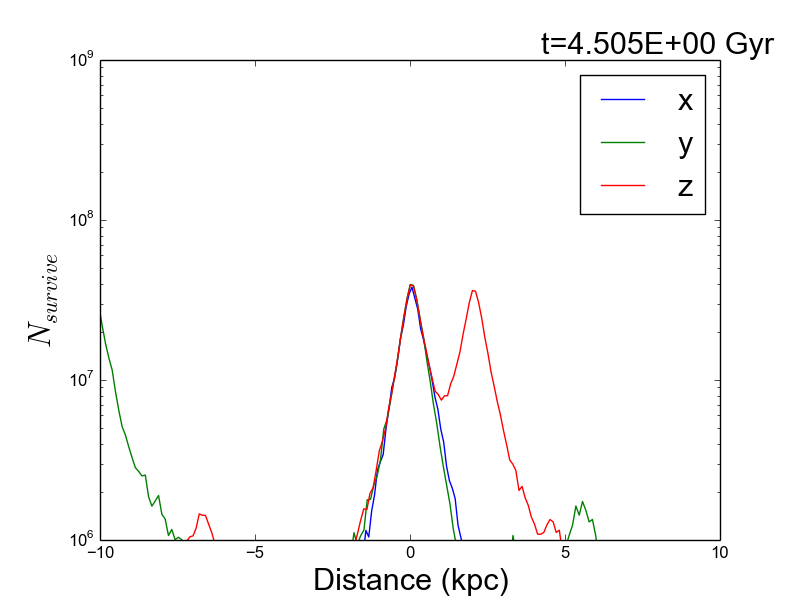} &
\includegraphics[scale=0.45]{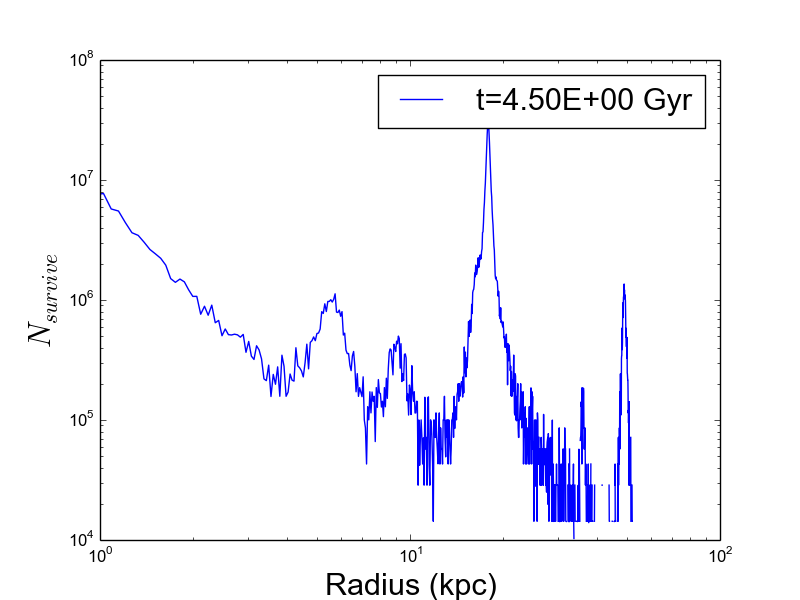} \\
\includegraphics[scale=0.45]{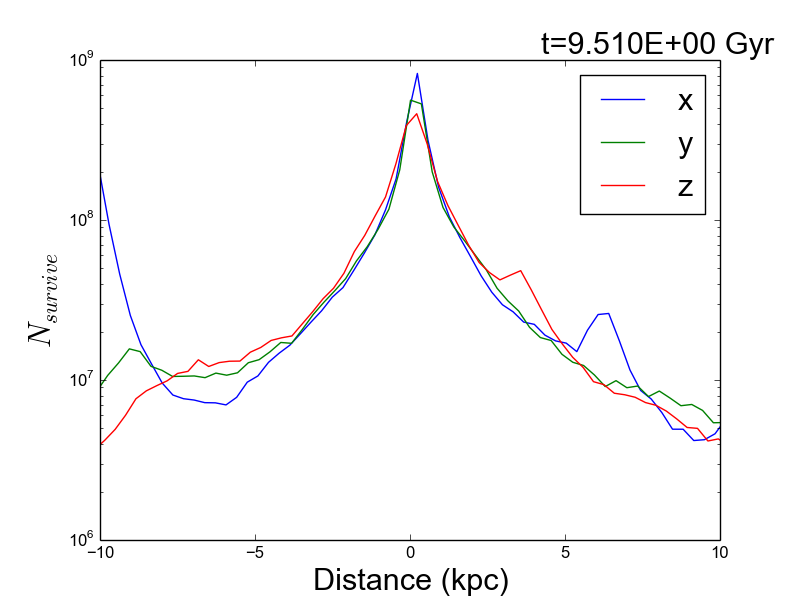} &
\includegraphics[scale=0.45]{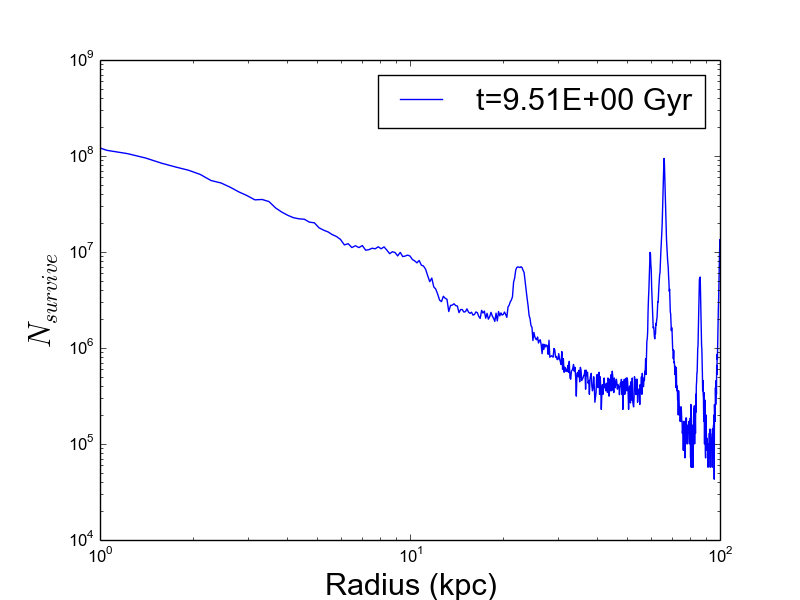} \\
\includegraphics[scale=0.45]{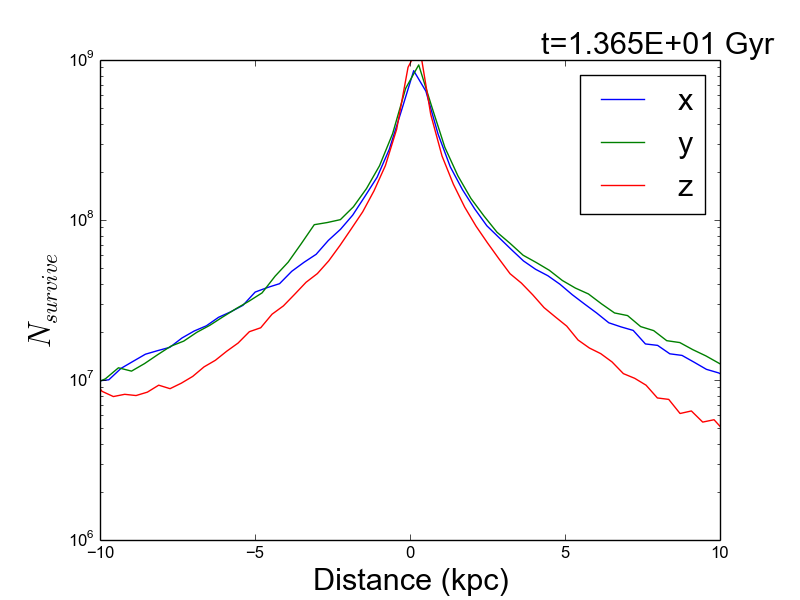} &
\includegraphics[scale=0.45]{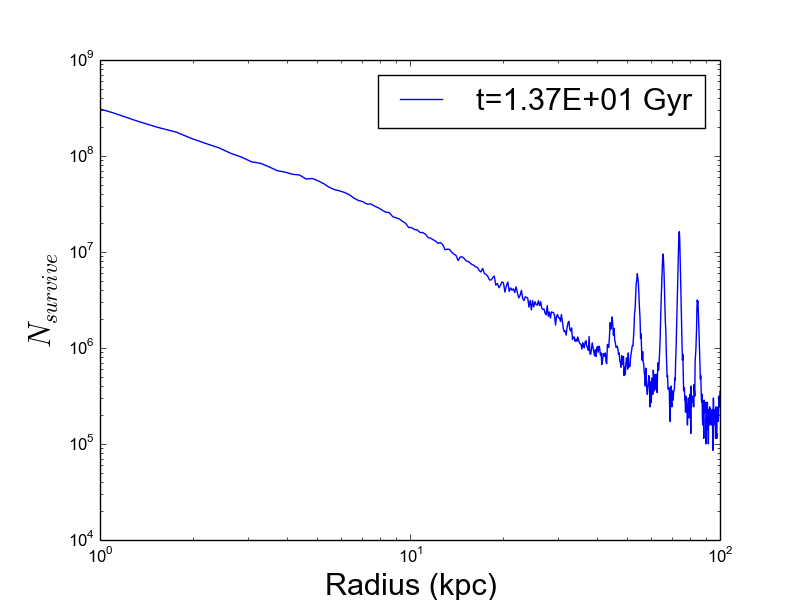} \\
\end{array}$
\caption{As Figure \ref{fig:M33_axial_nsurvive}, but for the Milky Way. \label{fig:MW_axial_nsurvive}}
\end{center}
\end{figure*}

Figure \ref{fig:MW_axial_nsurvive} shows the radial and axial profiles of this galaxy.  The larger number of satellites for this galaxy produce larger asymmetries, particularly in the axial profiles.  There is much less flattening of habitability in the z direction than in the case of M33, and the radial profiles follow a more uniform trend - there is no ``bump'' in habitability in the 4 to 10 kpc range as seen in M33.  Once again, we assume that the inner 1 kpc of this system, while recording a large number of surviving planets by our analysis, should be discounted due to the effect of the galaxy's central supermassive black hole.

\section{Discussion}

\subsection{General Trends}

Several features are present in both galaxies:

\begin{enumerate}
\item The main galaxy produces relatively broad GHZs,
\item The satellites also produce broad GHZs,
\item Streams of stars between satellites and galaxies produce relatively large numbers of habitable planets.  The satellites benefit from low supernovae rates, old stellar populations, less major mergers and the weak metallicity dependence of terrestrial planet formation.
\item The inner regions of the galactic disc contain a large number of surviving worlds by virtue of high stellar density, but the outer regions produce higher survival probabilities for individual star particles.  
\end{enumerate}

\noindent These trends are in line with recent GHZ calculations, such as \citet{Gowanlock2011}, who also show improved habitability above the midplane and at galactocentric radii of order a few kpc, much lower than the canonical limits given by \citet{GHZ}.  Given that we use some of Gowanlock et al's prescriptions for planet formation, etc., this is perhaps not surprising.  However, their models do not account for tidal streams and satellites.  Given our fairly loose constraints on terrestrial planet formation, and reduced metallicity suppressing giant planet formation, these structures on the outer limits of galaxies appear to be good sites for habitable planets.  However, we caution that our model assumptions remain simplistic, and there may be other hazards in these regions that we do not simulate.

Table \ref{tab:percentile} gives the first and third quartiles of $N_{survive}$ as a function of radius (equivalently, the 25th and 75th percentiles), and this shows some important differences between the M33 and Milky Way simulations, where we have again ignored the inner 1 kpc.  The M33 quartiles are initially well separated, and draw closer together with time to give a range between 3 and 10 kpc.  The Milky Way shows a less uniform evolution, where the quartiles are initially very close together, on the disc's outskirts (due to vigorous star formation in the interior), and then draw very far apart at intermediate times before settling to values of approximately 2 and 13 kpc.

\begin{table}
\centering
  \caption{The first and third quartiles (Q1 and Q3) of $N_{survive}$ as a function of radius for both galaxies\label{tab:percentile}}
  \begin{tabular}{c || cc || cc}
  \hline
  \hline
   Time (Gyr) & M33 Q1 (kpc) &  M33 Q3 (kpc) & MW Q1 (kpc) &  MW Q3 (kpc)  \\
   \hline
4.505 & 1.6 & 13.4 & 16.7 & 18.1 \\
9.51 & 2.78 & 9.95 & 2.8 & 66.1 \\
13.65 & 2.7 & 10.05 & 1.9 & 13.4 \\
 \hline
  \hline
\end{tabular}
\end{table}

As with all GHZ studies, we must consider these results in the light of the assumptions made, and the physics still missing from the analysis, as we do in the following section.

\subsection{Limitations of our Results}

\noindent This work is proof in principle that habitability can be determined from cosmological simulations - however, the simplicity of the analysis (like many GHZ analyses) leaves the conclusions we draw open to uncertainty.

Much of our issues stem from the fact that these simulations were not designed with GHZ calculation in mind.  Perhaps most pressingly, the resolution of the simulations limit us from exploring habitability at scales below a few hundred parsecs.  To date, no cosmological simulations exist that can resolve both the formation of individual stars and the global evolution of galaxy groups.  Our work pushes at the limits of what is currently possible with state-of-the art algorithms and computational power, but we are optimistic that future simulations will be able to address this.

Given that most sterilising events take place within a few tens of parsecs from a planet, we cannot investigate how neighbouring star particles influence each other.  While this has allowed a certain expediency in our calculations, it prevents us from calculating sterilisation rates convincingly in dense stellar environments, especially once natal star clusters have dissolved \citep{MoyanoLoyola2015}.  It also prevents us from measuring the effect of stellar encounters on local habitability \citep{Jimenez-Torres2013}, which is impossible without better estimates of the kinematics of individual stars.  Semi-analytic prescriptions for such properties could be employed, but we leave this for future work.

Our resolution limits also limit our ability to model individual sterilisation events.  Each star particle's star formation rate is calculated from semi-analytic approximations to reproduce empirical relations for star formation, in particular the Kennicutt-Schmidt relation \citep{kennicutt1983, kennicutt1998}.  We estimate supernova rates based on these star formation rates, assuming no time lag between star birth and star death.  However, Type I supernovae progenitors are low mass stars with a relatively long lifetime, and as such we should expect supernova rates that are less intense at early times than we calculate, with more supernova activity at later times as lower mass stars begin to ignite.  We should therefore note that our simulations underestimate habitability for young star particle ages, and overestimate it at older ages.  On the other hand, our prescriptions for Type I and Type II SNe tend to underproduce Type I relative to Type II, which goes some way to correcting this issue.  That being the case, this time lag issue is not the principal uncertainty in our habitability calculation, it being superseded by our uncertainty about the required radiation dose for sterilisation, as we discuss later.

The cosmological simulations do track the metallicity of the gas assuming this supernova activity is occurring, but the individual abundances of chemical elements are not traced.  This simplification is advantageous when attempting to determine broad trends in galaxy formation, but it leaves us unable to make more nuanced predictions about regions of favourable chemical composition.  For example, the abundance of iron and oxygen (and to some extent silicon) may give strong indications where there are large reservoirs of material to build planets with similar bulk compositions to the Earth, but a biota (at least based on what we know of terrestrial life) requires the presence of carbon, hydrogen, nitrogen, oxygen, phosphorus and sulfur (NCHOPS) as a mimimal set of elements colocalised at the scale of the organism and co-located with liquid water and an energy supply. These requirements cannot be easily predicted for any given planet.

In a similar vein, we do not consider the ``metals'' as a separate entity to the gas in the simulations.  While there are many density and pressure regimes where we expect volatile elements and compounds (`gas') to entrain the refractory materials (`dust') on similar dynamical trajectories, there are equally regimes where we expect these two phases to separate.  This separation is clearly visible in optical observations of the ``dust lanes'' of galaxies at various wavelengths, where dust obscures our view of the interior gas.  This evolution will affect the local metallicity, but it also affects the optical depth of the medium to biohazardous radiation.  One could imagine circumstances where planetary systems in regions of enhanced dust density are afforded a shield against ionising radiation and cosmic rays.  Equally, this biohazardous radiation may become biogenic radiation if it encounters dust/gas mixtures amenable to grain surface chemistry, enhancing the production of organic molecules and potentially even prebiotics \citep{Hill2003,Stark2014}.  This will have significant effects on the calculated habitability of low density regions such as tidal streams and the outer edges of satellite galaxies, and shows that a fully matured GHZ calculation must account for the anisotropy of radiative feedback as it encounters asymmetric distributions of intervening material.

On the subject of strong radiation sources, while we have considered the effect of Active Galactic Nuclei (AGN) implicitly by excluding the inner kiloparsec from our considerations of habitability, we do not explicitly calculate the interaction between the AGN and its environs.  AGN go through quiet and active phases depending on the black hole's accretion rate and the galaxy's own accretion history, and we expect their effects to extend well beyond the 500 pc radius of our star particles.  Indeed, it is thought that ``AGN feedback'' may be an important mechanism for quenching star formation \citep{Springel2005a,DiMatteo2005}, although it might also have the opposite effect \citep{Ishibashi2013}.  If the accretion rate of the central supermassive black hole - and consequently the fuelling of the central engine of the AGN - is included in the simulation, then the models could link mergers and accretion to AGN activity and subsequent star formation, which has consequences both for galaxy evolution and galactic habitability.

Another powerful source of sterilising radiation comes from gamma ray bursts (GRBs), powerful explosions generated by the mergers of compact objects, or the collapse of extremely massive stars  (see \citealt{Fishman1995,Berger2014} for reviews).  As GRBs tend to be more frequent in the past than in the present, and their sterilisation radius is large compared to the Galactic disc, the GHZ at early times could be even more asymmetric and patchy, if it exists at all.  Recent studies have characterised the GRB threat to Earth by studying the GRB luminosity function \citep{Piran2014}, which confirms the trend that GRBs are more frequent in low metallicity environments.  As a result, the GRB rate is higher at earlier times, which suggests that the Milky Way and most other galaxies might have very little GHZ to speak of before $z=0.5$.  This sort of ``global regulation mechanism'' has been proposed to answer Fermi's Paradox, or the observed ``Great Silence'' of transmissions from, or other evidence of, intelligent technological civilisations \citep{Annis,Vukotic_and_Cirkovic_07}.  More modelling in this area is essential.  

Perhaps most importantly, this work has demonstrated the current limitations of the GHZ concept.  Modelling biosphere formation using a single parameter, metallicity, is unlikely to yield the ``ground truth'' of how habitable worlds are formed.  Our understanding of the timescales required for metazoan lifeforms to evolve remains informed by Earth's geological history, but essentially incomplete due to gaps in our knowledge of comparative exoplanetology, for example, our understanding of plate tectonics on super Earths, which remains a source of debate \citep{Stamenkovic2010,Stamenkovic2014}.  We also remain quite ignorant of the levels of hazardous radiation required to definitively exterminate a biosphere, evidenced by the fact that the only known biosphere - our own - continues to persist.  A significant fraction of damaging cosmic rays are absorbed or scattered away from the Earth by the Sun's heliosheath, and the Earth's own protective magnetic field.  Simply stating that a SN is "deadly within 8pc" is an extremely glib statement, and belies the subtleties involved in determining the effect of hazardous interstellar radiation on planets within stellar magnetospheres \citep{Gehrels2003,Martin2009, Thomas2015}.  Future GHZ modelling must address these concerns, first put forward by \citet{Prantzos2007}, and which remain unresolved.

Our principal contribution to this debate has been to demonstrate that the GHZ will be morphologically complex.  We have used state-of-the-art simulations to reproduce galaxy assembly histories, but have been forced to use relatively simple chemical evolution modelling and planet formation criteria to leverage the simulation data in ways that it wasn't designed.  Future work in this area must use better chemodynamical models to better trace the distribution of NCHOPS elements, as well as the stellar population (see e.g. Gibson et al., submitted), while still resolving the assembly history of the Local Group accurately in cosmological context.

\section{Conclusions }\label{sec:conclusions}

\noindent For the first time, we have estimated galactic habitability in three dimensions by analysing high resolution numerical simulations of the formation of the Local Group.  We apply habitability metrics commonly found in other studies of 1D azimuthally symmetric Galactic Habitable Zones (GHZs).

Thanks to mergers and the accretion of satellite galaxies, we find that if the GHZ exists, it must be fundamentally asymmetric.  These dynamical events produce tidal streams and spiral arms which produce elliptical and even triaxial regions of improved habitability.  While the simulations indicate that large numbers of habitable, non-irradiated planets can be found in the inner regions of galaxies and their satellites, the probability that an individual planetary system contains habitable worlds increases with distance from the galactic centre, with its peak near the edges of galactic discs.  Regions above the midplane are also favoured locations, but the density of stars in these areas reduces the total number of habitable planets.

However, the idea that the outer edges of galactic discs are ``more habitable'' is not wholly confirmed by our analysis (see e.g. Table \ref{tab:percentile}).  For the Milky Way, this appears to be true at early times, where the first and third quartiles of all habitable planets are initially at large distances, but over time this interquartile range increases greatly before settling to values of approximately 2 and 13 kpc.  For M33, we see quite opposite evolution, but a similar final interquartile range between 3 and 10 kpc.  These large changes suggests that \citet{Prantzos2007}'s criticisms are vindicated - we do not see a GHZ that is well constrained in space/time, and habitable planet formation appears to be possible at many locations in a galaxy.  Furthermore, the divergence in evolution between both galaxies suggests that assembly history plays a crucial role in subsequent habitability.

However, we acknowledge that galactic habitability is still maturing as a conceptual tool.  This paper has demonstrated that cosmological simulations can be extremely informative regarding both galactic habitability and the weaknesses of our assumptions about planet formation and sterilisation.  As these simulations continue to improve in spatial resolution and input physics, we believe that the galactic habitable zone will become as useful as the circumstellar habitable zone in assessing the astrobiological potential of our local cosmic volume.

\section*{Acknowledgments}

DF acknowledges support from STFC consolidated grant ST/J001422/1 , and the "ECOGAL" ERC Advanced Grant.  PD
acknowledges the support of the Addison Wheeler Fellowship awarded by the Institute of Advanced Study at Durham University.  NIL is supported by the Deutsche Forschungs Gemeinschaft (DFG).  The authors are grateful to the reviewers, whose comments and suggestions greatly improved this manuscript.

\bibliographystyle{mn2e} 
\bibliography{GHZ_sim}

\appendix

\label{lastpage}

\end{document}